\documentclass[hyper(*)]{JHEP3} 

\usepackage{epsfig,multicol,bbm}

\newcommand\fverb{\setbox\fverbbox=\hbox\bgroup\verb}
\newcommand\fverbdo{\egroup\medskip\noindent%
            \fbox{\unhbox\fverbbox}\ }
\newcommand\fverbit{\egroup\item[\fbox{\unhbox\fverbbox}]}
\newbox\fverbbox

\title{$N=2$ SU Quiver with USP Ends or SU Ends with Antisymmetric Matter}

\author{Dimitri Nanopoulos $^{1,2,3,a}$, and Dan Xie $^{1,b}$

\\$^{1}$George P. nd Cynthia W.Mitchell Institute for Fundamental Physics,
Texas A\&M University, College Station, TX 77843, \\
$^{2}$Astroparticle physics Group, Houston Advanced Research
Center (HARC), Mitchell Campus, Woodlands, TX 77381, USA\\
$^{3}$Academy of Athens, Division of Nature Sciences, 28
panepistimiou Avenue, Athens 10679, Greece\\
E-mail: \email{$^{a}$dimitri@physics.tamu.edu,
$^{b}$fogman@neo.tamu.edu}}

\preprint{ACT-08-09, MIFP-09-31}  

\abstract{We consider the four dimensional scale invariant $N=2$ SU
quiver gauge theories with $USp(2N)$ ends or $SU(2N)$ ends with
antisymmetric matter representations. We argue that these theories
are realized as six dimensional $A_{2N-1}$ $(0,2)$ theories
compactified on spheres with punctures. With this realization, we
can study various strongly coupled cusps in moduli space and find the
S-dual theories. We find a class of isolated superconformal field
theories with only odd dimensional operators $D(\phi)\geq3$ and
superconformal field theories with only even dimensional operators
$D(\phi)\geq4$.}


\dedicated{}

\begin{document}


\section{Introduction}
Recently, Gaiotto \cite{Gaiotto} proposed a remarkable method to
describe four dimensional scale invariant $N=2$ quiver gauge
theories with bi-fundamental and fundamental fields. Four
dimensional $N=2$ quiver gauge theories are defined as the
compactification of six dimensional $A_N$ $(0,2)$ theories on
Riemann surfaces with punctures. The marginal couplings of the
quiver gauge theory are determined as the moduli of punctured
Riemann surfaces. The punctures are labeled by Young Tableaux and
there is a correspondence between puncture type with the flavor
symmetry of the four dimensional quiver gauge theory. The various
S-dual frames were shown to correspond to different degeneration
limits of this punctured Riemann surface. This generalized the
previous observation of Argyres-Seiberg dualtiy \cite{Argy} on the
infinite strongly-coupled region of SU(3) gauge theory with six
fundamental hypermultiplet matters: they found a S-dual theory in
which a weakly coupled $SU(2)$ theory coupled with a isolated
interacting $E_6$ superconformal field theory in which a $SU(2)$
subgroup is gauged. Gaiotto's construction generalized their result
and is used to find a new family of isolated superconformal field
theories with $SU(N)^3$ flavor symmetry \cite{Gaiotto}.

$D=6$ is the maximal dimension in which we can formulate a
superconformal field theory. The six dimensional  $(0,2)$
superconformal field theory has the famous ADE classification. The
compactification of six dimensional theory on a Riemann surface
provides a lot of insights on four dimensional conformal field
theory \cite{witten1}.  For instance, if we compactify $A_{N-1}$
theory on a smooth torus, the $SL(2,Z)$ invariance of the four
dimensional $ N=4$ SU(N) gauge theory is directly related to the
$SL(2,Z)$ modular group of the torus. Gaiotto's construction
provides another six dimensional framework to understand S-dualities
of four dimensional $N=2$ scale invariant theory. Here we allow
codimension two defects \cite{witten2} of this $A_{N-1}$ theory.
These defects have singularities at the punctures from which we can
also read the flavor symmetries of four dimensional theory.

It is definitely interesting to extend this analysis to other four
dimensional $N=2$ scale invariant theories. It is the main aim of
this note to extend this analysis to the $N=2$ SU linear quiver
gauge theories with $USp(2n)$ group on the end or with $SU(2n)$
group on the end with antisymmetric matter representation. These
theories have a Type IIA brane construction involving  two $O6$
orientifold planes. In type IIA theory, we have a $NS5-D4$ system in
the background of $O6$ planes and D6 branes. The $NS5-D4$ system
lifts to a single M5 brane wrapped on a smooth Riemann surface in an
M theory background describing the M theory lift of O6 planes and D6
branes. The Riemann surface is identified with the Seiberg-Witten
curve. We can rewrite the Seiberg-Witten curve in a way along with
Gaiotto's construction on an ordinary $SU$ quiver. It can be shown
that these theories can be realized as the compactification of
$A_{k-1}$ theory on spheres with punctures. In particular, we
confirm that SU quiver gauge theory with $USp$ ends falls in the
same duality web as the quiver gauge theory with pure SU nodes
\cite{Gaiotto}. We also identify the dual quiver to the theory with
SU ends with antisymmetric representations. We will study the
infinite strongly coupled region of some theories and we can see the
emergent weakly coupled node coupled to an isolated $E_6$ or $E_7$
superconformal theory \cite{min1,min2}. We also find an family of
isolated superconformal field theories with only odd dimensional
operators $D(\phi)\geq3$ and superconformal field theories with even
dimensional operators $D(\phi)\geq4$.

This note is organized as follows: in section 2, we review Gaiotto's
construction on $A_k$ theory; In section 3, we describe the brane
construction of our model and rewrite the Seiberg-Witten curve in a
form which makes the description of compactification of the six
dimensional $(0,2)$ theory manifest; In section 4, we describe
explicitly the six dimensional description of some specific examples. In
section 5, we study the various degeneration limits of our theories.
Finally, we give our conclusion.

\section{Review of $(0,2)$ $A_{k-1}$ Theory on Punctured Riemann Surfaces}
We consider a four dimensional $N=2$ linear quiver gauge theory with
a chain of SU groups
\begin{equation}
SU(n_1)\times SU(n_2)\times...\times SU(n_{n-1})\times SU(n_n),
\end{equation}
and bifundamental hypermultiplets between the adjacent gauge groups
and $k_a$ extra fundamental hypermultiplets for $SU(n_a)$ to make
the gauge couplings marginal. The marginality of gauge couplings
imposes the constraints on the number of fundamentals:
\begin{equation}
k_a=(n_a-n_{a-1})-(n_{a+1}-n_a),
\end{equation}
we define $n_0=0, n_{n+1}=0$. Since $k_a$ is nonnegative, we have
\begin{equation}
n_1<n_2<...n_r=..n_l>n_{l+1}...n_n.
\end{equation}
Let's first consider the left tail. Let us denote $N=n_r=...=n_l$,
so that the non-increasing number $(n_a-n_{a-1}), a\leq r$ satisfies
the relation $\sum_{a=1}^{a=r}(n_a-n_{a-1})=N$. For the right tail,
the non-decreasing number $n_{a}-n_{a+1}$ starting from $a=n$ also
satisfies the relation $\sum_{a=l}^{a=n}(n_{a}-n_{a+1})=N$. So we
associate a Young Tableaux with total boxes N for each tail (see
Figure 1a). The flavor symmetry of this linear quiver is
$U(1)^{n+1}\times \sum_{a=1}^r SU(k_a)\times \sum_{a=l}^n SU(k_a)$,
which can be read explicitly from the quiver diagram.
\begin{figure}
\begin{center}
\includegraphics[width=4in, height=2.5in]
{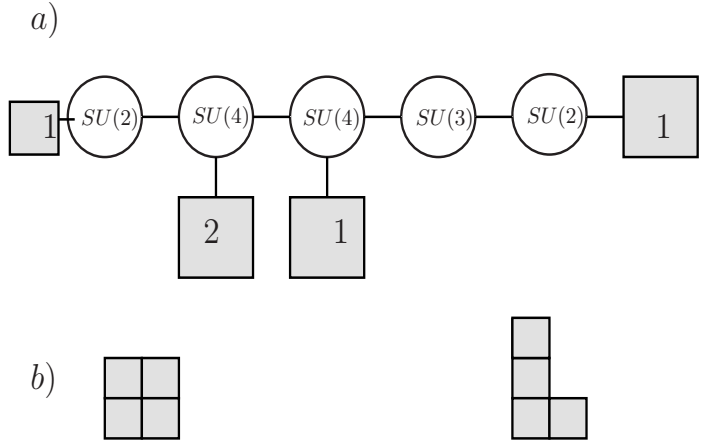}
\end{center}
\caption{a) A $N=2$ linear quiver with $N=4$; b) The Young Tableaux
associated with left and right tail. }
\end{figure}

\begin{figure}
\begin{center}
\includegraphics[width=4in, height=1.5in]
{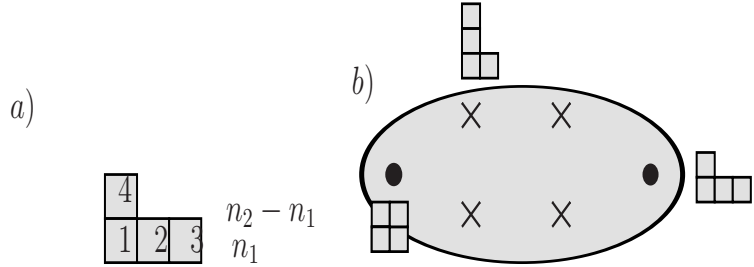}
\end{center}
\caption{a) Young Tableaux associated with the tail in a linear
quiver gauge theory with $N=4$, $p_1=1-1=0$, $p_2=2-1=1$,
$p_3=3-1=2$, $p_4=4-2=2$, the flavor symmetry is $SU(2)$; b) The
punctured sphere for $(0,2)$ $A_3$ theory compactification, each
puncture is labeled by a Young-tableaux }
\end{figure}

The Seiberg-Witten curve for this theory has been solved in
\cite{witten3}. The curve is rewritten in the following form
\cite{Gaiotto}:
\begin{equation}
t^N+\sum_{i=2}^N\phi_i(x)t^{N-i}=0,
\end{equation}
where $x$ is the coordinate on a sphere; and the Seiberg-Witten
differential is simply $\lambda=tdx$.

$\phi_i(x)dx^i$ are degree i differentials on the sphere with poles
at $n+3$ punctures, say with $x=0, \infty$ and $x_1...x_{n+1}$. The poles
at $x_1...x_{n+1}$ are of order $p_i=1$, and are called basic
punctures; the pole at $t=0$ has order $p_i=i-s$, where $s$ is the
height of the Young Tableaux we have just constructed. The pole at
$t=\infty$ can be similarly determined from the Young Tableaux
associated to the other tail.

This motivates the following six dimensional description of these
four dimensional $N=2$ superconformal field theories: the linear
theory is realized as a six dimensional $A_{N-1}$ theory
compactified on a sphere with $n+3$ punctures, and the punctures are
labeled by the Young Tableaux (see Figure 1b). There are also
defects at the punctures. Recall that six dimensional $A_{N-1}$
theory has operators of dimension $2,3,...,N$. Compactification on
Riemann surface involves the ordinary twisting to preserve
supersymmetry, this twisting turns the dimension $i$ operators into
a degree $i$ meromorphic differential $\phi_idx^i$ on Riemann
surface.

The orders of the poles are determined from the Young Tableaux by
using the formula $p_i=i-s$, where $i$ is the label of the $i$th box
and s is the height of $i$th box in the Young tableaux (see Figure
2a). The dimension of the space of these meromorphic differentials
is given by
\begin{equation}
 dimension~of~\phi_i= \sum_{punctures~d=1}^{n+3}p^{(i)}_d+1-2i
\end{equation}
The parameters of these differential are identified with dimension
$i$ operators of the four dimensional theory, i.e. the parameters
for the Coulomb branch. The Seiberg-Witten curves describing the low
energy effective theory of these models are also expressed in terms
of these operators:
\begin{equation}
t^N+\sum_{i=2}^N\phi_i(x)t^{N-i}=0
\end{equation}

The gauge coupling constants are identified with the moduli of the
punctured sphere $M$. The moduli space for a sphere with $n+3$
punctures is $n$ dimensional which is identified with the n coupling
constants of our linear quiver. The duality group is $\pi_1(M)$.

We can also determine the flavor symmetry from the punctures. This
can be determined by studying the mass-deformed theory. The
Seiberg-Witten curve for mass-deformed theory has the additional
term $\phi_1t^{N-1}$; we can do a linear transformation on $t$ to
eliminate this linear term. Additionally, the Seiberg Witten curve
is changed to $\lambda=t^{'}dx$, where $t^{'}$ is the new variable.
The change of Seiberg Witten curve only redefines the mass
parameters. The mass parameters are now identified as the residue of
this new Seiberg-Witten differential at the puncture. The pattern of
the residue is in one-to-one correspondence with the Young Tableaux
of this puncture. For each column of Young Tableaux with height
$l_h$, there are $l_h$ same residue for Seiberg Witten differential.
The flavor symmetry of this puncture is
\begin{equation}
S(\prod_{l_h>0}U(l_h)).
\end{equation}
(See Figure 2a for an example). One can check that this
characterization of flavor symmetry matches that read from the
linear quiver gauge theory. If the Young Tableaux of a puncture has
only one row and the flavor symmetry is then $SU(N)$, we call the
puncture a full puncture; if the Young Tableaux has two columns, one
of them has $N-1$ boxes, the other has one box, and the flavor
symmetry is $U(1)$, then we call this a basic puncture. The punctures
associated with the tails are called generic punctures.

This then gives a complete description of four dimensional $N=2$
quivers from six dimensional point of view. One of the great virtues
of this description is that we can derive various weakly coupled
descriptions of these superconformal field theories by studying
various degeneration limits of the punctured sphere.
\begin{figure}
\begin{center}
\includegraphics[width=4in,]
{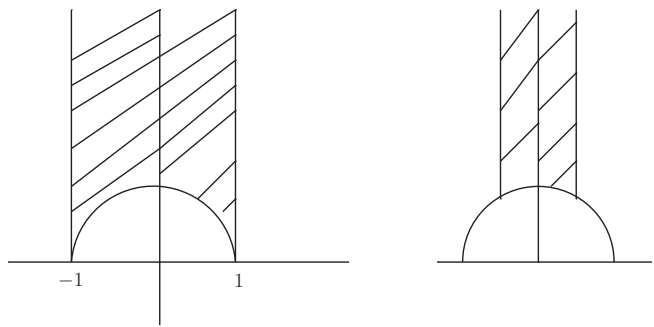}
\end{center}
\caption{a) The fundamental domain of $H\over \Gamma(2)$. Here $H$
is the upper half plane, $\Gamma(2)$ is the duality group of a
sphere with four punctures; b)The fundamental domain of $H/
SL(2,z)$. }
\end{figure}

We first study the simplest case of four dimensional scale invariant
SU(2) theory with four fundamental matter hypermultiplets. The full
flavor symmetry is $SO(8)$, we can write it in a form with only
manifest $SU(2)_a\times SU(2)_b\times SU(2)_c\times SU(2)_d$ flavor
symmetry. This description will make the six dimensional
interpretation manifest.

It is shown in \cite{witten4} that the duality group of this theory
is $SL(2,z)$. The duality group is the combination of $\Gamma(2)$
(the symmetry group of a sphere with four punctures) and the
triality of $SO(8)$ flavor symmetry. The triality of  $SO(8)$ flavor
symmetry permutates four manifest $SU(2)$ flavor groups.
\begin{figure}
\begin{center}
\includegraphics[width=4in,]
{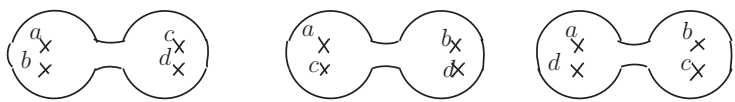}
\end{center}
\caption{The various weakly coupled limit of SU(2) theory with four
fundamental matter. The narrow strip denotes the weakly coupled
SU(2) gauge group. The punctures are associated with flavor symmetry
$SU(2)$.}
\end{figure}

Before we use the triality symmetry, let us consider the moduli
space which is shown in Figure 3a). The six dimensional description
is shown in Figure 4. We have three different weakly coupled
descriptions as the different degeneration limit of the punctured
sphere. These three weakly coupled descriptions correspond to the
three cusps $(1,0), (-1,0), \infty$ in the moduli space in Figure
3a). After using the permutation symmetry, the four punctures are
identical and the duality group is enhanced to $SL(2,z)$. The moduli
space becomes $H/SL(2,z)$, which is shown in Figure 3b). The three
weakly coupled descriptions are identical and we have only one
weakly coupled description, which corresponds to the only cusp
$\infty$ in the new moduli space.

Next we study the $SU(3)$ theory with six fundamentals. This is the case
considered by Argyres and Seiberg \cite{Argy}. The moduli space of
this theory is depicted in Figure 3a). The manifest flavor symmetry
is $U(1)^2\times SU(3)^2$ in our description. The two different
degeneration limits of six dimensional description are depicted in
Figure 5. Figure 5a) is ordinary description in which the $SU(3)$
gauge coupling can be made arbitrarily weak.
\begin{figure}
\begin{center}
\includegraphics[width=4in,]
{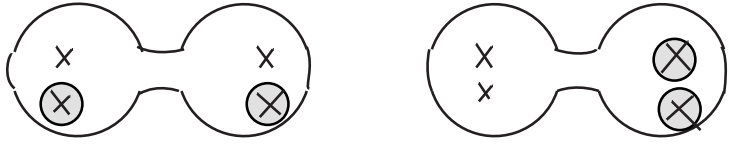}
\end{center}
\caption{ a) The weakly coupled SU(3) description, here the cross
denotes $U(1)$ puncture and circle cross denotes $SU(3)$ puncture;
b) Description with weakly coupled SU(2) gauge group }
\end{figure}
The description shown in Figure 5b) is rather surprising, since
this description corresponds to the infinitely strongly coupled
region of the description Figure 5a). However, it is quite natural from
the six dimensional point of view, it is just one degeneration limit
of the punctured Riemann surface. This gives a complete picture of
various weakly coupled corner in the moduli space, since these are
the only cusps in the moduli space shown in Figure 3a)(two of the cusps
are identical).

Actually, we can determine which gauge group is becoming weakly
coupled and what kind of puncture is left when we completely turn
off this gauge group. When a basic puncture collides with a generic
puncture, then the gauge group becoming arbitrarily weak is
$SU(n_1)$, where $n_1$ is the first row of the Young Tableaux
associated with the generic puncture. When we completely turn off
this weakly coupled gauge group, we leave a puncture with a Young
Tableaux whose first row has  $n_2$ boxes, while the other rows are
not changed. In our particular case, the generic puncture is a
simple puncture, with $n_1=2$, $n_2=3$, after decoupling, we create
a new puncture with only one row $n_2=3$. The flavor symmetry is
$SU(3)$ according our rules. This is shown in Figure 6.
\begin{figure}
\begin{center}
\includegraphics[width=4in,]
{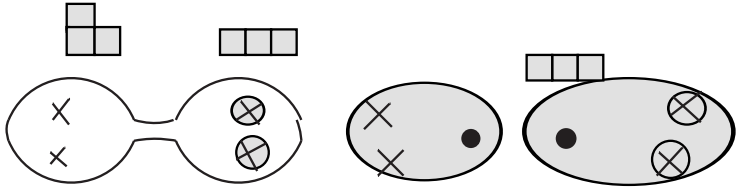}
\end{center}
\caption{ The collision of two basic puncture create a new puncture
 when we turn off the weakly coupled gauge group, we also draw the Young Tableaux associated with this new created puncture, in our
 particular case, the flavor symmetry is $SU(3)$.}
\end{figure}
The theory associated with three punctures with $SU(3)$ gauge group
is an interacting superconformal field theory. For a sphere with
three punctures, there is no moduli so this theory is an isolated
fixed point. As we described earlier, the Seiberg Witten curve is
\begin{equation}
t^3+{U\over(x-1)^2x^2}=0
\end{equation}
$U$ is the dimension 3 operator of this theory. The manifest flavor
symmetry is $SU(3)^3$, which is enhanced to $E_6$. This is the
famous $N=2$ $E_6$ superconformal field theory. The S-dual theory of
$SU(3)$ theory is recovered easily in this way. It is a $SU(2)$
theory with one fundamental and coupled with $E_6$ superconformal
field theory, and a $SU(2)$ subgroup of $E_6$ is gauged and
identified with the weakly coupled $SU(2)$ gauge group. The number
of fundamentals can be either determined from the three punctured
sphere, or from the counting of the conformal anomaly of the $SU(2)$
theory: the $SU(3)$ puncture provides the conformal anomaly equal to
three fundamental of $SU(2)$, so we need one extra fundamental to
compensate the conformal anomaly.

Similarly, we can also construct isolated superconformal theories
with flavor symmetry $SU(N)^3$ by considering the degeneration limit
of superconformal quiver $SU(2)\times SU(3)...SU(N)^{N-2}\times
SU(N-1)...SU(2)$\cite{Gaiotto}. We have a total of $3N-3$ basic
punctures on the sphere. It is easy to deduce that when $N-1$ basic
punctures collide (we can collide basic punctures step by step), a
$SU(N-1)$ gauge group becomes weakly coupled and if we turn off this
gauge coupling we are left a $SU(N)$ puncture. In the core of the
degenerated sphere, we have a theory with three $SU(N)$ punctures.
See Figure 7. The supergravity dual of this $T_N$ theory is found in
\cite{Gaiotto2}.
\begin{figure}
\begin{center}
\includegraphics[width=4in,]
{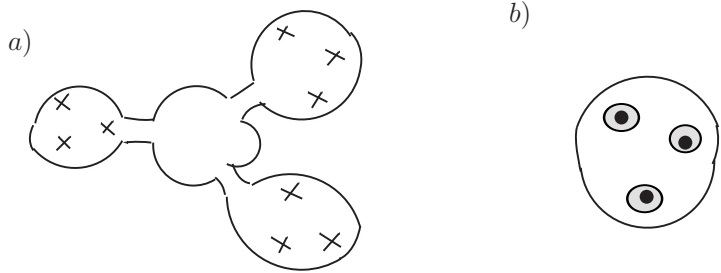}
\end{center}
\caption{ a) The collision of three groups of $N-1$ basic punctures;
b) $T_N$ theory with three full punctures with $SU(N)$ flavor
symmetry}
\end{figure}

We have known how a basic puncture collide with a generic puncture.
It is interesting to consider the collision of two generic
punctures. It is shown in \cite{Gaiotto} that when two generic
$SU(2)$ punctures associated with a Young Tableaux with two columns
of equal heights collide, an USp gauge group becomes weakly coupled.
We will check in this paper that this conjecture is correct. We will
also consider the collision of a special $U(1)$ puncture and a
$SU(2)$ puncture.

The above construction can be generalized to four dimensional $SU$
quiver with loops \cite{Gaiotto}. It can also be generalized to
$D_N$ theory \cite{yuj1} and there is also brane web construction in
\cite{yuj2}. There is an interesting relation between four
dimensional $N=2$ gauge theories and liouville correlation function
\cite{Gaiotto3}.

\section{$N=2$ SU Quiver with USp Ends or SU Ends with Antisymmetric matter}
Four dimensional $N=2$ superconformal $SU$ field theory with USp
ends or SU ends with antisymmetric representations can be derived by
adding orientifold six planes to Type IIA D4-NS5 brane system
\cite{uranga}. The solution of the model \cite{argy2} can be found
after lifting the above brane configuration to M theory along the
similar line as in \cite{witten3}

We first consider D4 and NS5 branes system in type IIA theory; We
also include two orientifold six planes and 8 D6 branes so that the
net RR charges cancels. The k four branes lie along the directions
$x_0,x_1,x_2,x_3, x_6$; we take $x^6$ coordinate compact. The NS5
branes lie along $x^0,x^1,x^2,x^3,x^4,x^5$ directions. The
orientifold six planes extend along $x_0,x_1,x_2,x_3,x_7,x_8x_9$
directions. It corresponds to the space time transformation
\begin{equation}
h: (x_4,x_5,x_6)\rightarrow (-x_4,-x_5,-x_6),
\end{equation}
together with the world sheet parity $\Omega$ and $(-1)^{F_L}$. The
D6 branes are parallel to $O6$ planes.

There are three main families of $N=2$ quiver gauge theory with
these brane configurations, depending on the positions of the NS
branes:

\begin{figure}
\begin{center}
\includegraphics[width=4in,]
{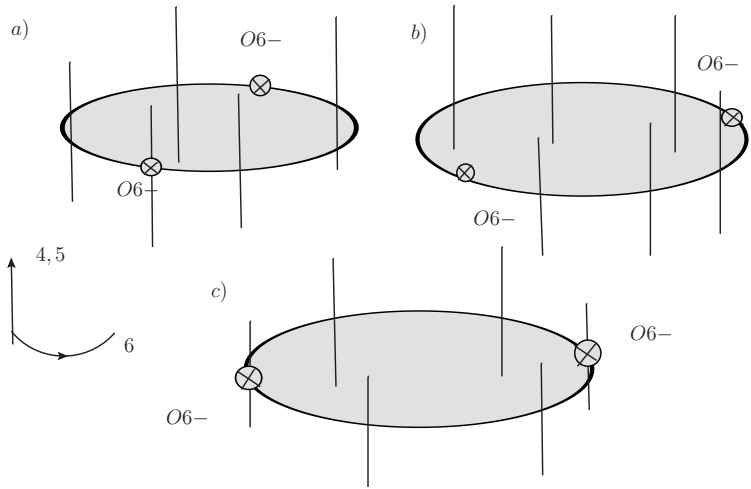}
\end{center}
\caption{The tree families of brane configurations in the background
of two negatively charged O6-planes. The short vertical lines
represent the NS branes, the crossed circles are the orientifold
planes. The D6 branes is put in between the NS branes, we omit them
in the picture. }
\end{figure}

i) The number of NS branes is odd, $N=2r+1$. Only one NS5 brane
intersect with the orientifold plane. One typical brane
configuration is depicted in Figure 8a. The quiver gauge theory is
\begin{equation}
USp(k)\times SU(k)^{r-1}\times SU(k),
\end{equation}
$k$ must be even since for USp group the rank must be even. We have
the bifundamental matter fields between the adjacent group.  Two
fundamentals are attached at the USp node and we have two
fundamentals and one antisymmetric hypermultiplet at the last
$SU(k)$ node. The flavor symmetry is $SO(4)\times U(1)^r\times
SU(2)\times U(1)$. The $SO(4)$ flavor symmetry is from the two
fundamentals of $USp$ node, and the last $SU(2)\times U(1)$ is from
the two fundamentals of the $SU$ ends.

Note that the antisymmetric representation of $SU(k)$ is real, so
the flavor symmetry of this representation is $Usp(2)=SU(2)$. In
this paper, however, we do not consider the mass deformation of
antisymmetric matter, so we do not include the flavor symmetry
associated with it. Use the isomorphism $SO(4)=SU(2)\times SU(2)$,
the total flavor symmetry is $SU(2)\times SU(2)\times U(1)^r\times
SU(2)\times U(1)$.

ii)The number of NS branes is even. $N=2r$, and there are no NS
branes intersecting the O6-planes. One example is shown in Figure
8b. The quiver gauge theory is
\begin{equation}
USp(k)\times SU(k)^{r-1}\times USp(k).
\end{equation}
We have the bifundamentals between the adjacent group and two
fundamentals at the first and last $USp$ gauge factor. The flavor
symmetry is $SU(2)\times SU(2)\times U(1)^r\times SU(2)\times
SU(2)$.

iii)The number of NS branes is even $N=2r$. There are two NS branes
intersecting with the O6-planes. One configuration is shown in
Figure 1c. The quiver gauge theory is
\begin{equation}
SU(k)\times SU(k)^{r-1}\times SU(k)
\end{equation}
Besides the bifundamental matters, we have two fundamentals and one
antisymmetric at the first and the last $SU$ factor. The flavor
symmetry is $ U(1)\times SU(2)\times U(1)^r\times SU(2)\times U(1)$.

When $r=0$, the above theories are degenerate as

i) $USp(k)$ with a traceless-antisymmetric and 4 fundamentals.

ii) Also a $USp(k)$ with traceless-antisymmetric and 4 fundamentals,
this is only for the massless antisymmetric matter, the mass
deformation for this matter is not allowed.

iii) $SU(k)$ with 2 antisymmetric hypermultiplets and 4
fundamentals.

The Seiberg-Witten curves for those theories are derived by lifting
the Type IIA configuration to M theory \cite{argy2}. Here we briefly
review the derivation. The NS5-D4 brane configuration is lifted to a
single M5 brane wrapped on a Riemann surface in $O6-D6$ background.
In lifting to M theory, we grow a circular dimension $x_{10}$ with
radius $R$. Define the variables
\begin{equation}
v=x^4+ix^5,~~~ s=(x^{10}+ix^6)/(2\pi R)
\end{equation}
Before orbifolding, the background space is $\tilde{Q}=C\times T^2$.
The $Z_2$ identification of the orientifold is $(v,s)\simeq(-v,-s)$.
The M theory background is therefore the orbifold space
$Q=\tilde{Q}/Z_2$.

We only need the complex structure of this orbifold background. To
do this, we first write an algebraic equation of torus. The torus
can be written as an complex curve in the weighted projective space
$CP^2_{(1,1,2)}$. $CP^2_{(1,1,2)}$ is defined as the space
$(w,x,y)/(0,0,0)$ modulo the identification
\begin{equation}
(\lambda\omega,\lambda x,\lambda^2 \eta)\simeq
(\omega,x,y),~~~\lambda\in C^{*}.
\end{equation}
The torus is represented as
\begin{equation}
\eta^2=\prod_{i=1}^4(x-e_i\omega),
\end{equation}
where the numbers $e_i$ encode the complex structure $\tau$ of the
torus in usual way.

The $Z_2$ automorphism of the torus is $\eta\rightarrow -\eta$ with
$\omega$ and $x$ fixed. The $Z_2$ identification of the orientifold
background becomes $(v,\omega,x,\eta)\simeq(-v,\omega,x -\eta)$. The
fixed points are
\begin{equation}
(0,1,e_i,0)~~~~i=1,2,3,4,
\end{equation}
we write it in $\omega=1$ patch.

Let us define $Z_2$ invariant variables
\begin{equation}
y\equiv\eta v,~~~~~~z=v^2,
\end{equation}
the orbifolded background $Q$ (without mass deformation for the
fundamental matter) is
\begin{equation}
y^2=z\prod_{i=1}^4(x-e_i\omega).
\end{equation}
In the following, we write all the formulas in the patch $\omega=1$,
so the orbifold equation is
\begin{equation}
y^2=z\prod_{i=1}^4(x-e_i).
\end{equation}
The mass deformed (which corresponds to mass deformation to four
fundamental matters induced by D6 branes) background is
\begin{equation}
y^2=z\prod_{i=1}^4(x-e_i)+Q(x)
\end{equation}
and
\begin{equation}
Q(x)=\sum_{j=1}^4\mu_j^2\prod_{k\neq j}[(x-e_k)(e_j-e_k)]
\end{equation}

The Seiberg Witten curve for those field theories is a Riemann
surface embedded into above background. We can first write the
Seiberg Witten curve for the brane configuration before orbifolding,
which is just the elliptic model in \cite{witten3}, and then require
the curve invariant under the $Z_2$ transformation. For the elliptic
model, the bifundamental masses satisfy the relation
$\sum_{\alpha}m_\alpha=0$, so to get the most generic mass-deformed
theory, the background is not simply $C\times T_2$ but an affine
model. There is no such problem for our model; before orbifolding,
the relation $\sum_{\alpha}m_\alpha$ still applies, however, after
orbifolding, the bi-fundamental masses are all independent(the
orbifold images of D4 branes have opposite $v$ coordinates, so the
bi-fundamental mass for two images are opposite). We do not need to
change the background to an affine bundle to allow most generic mass
deformation for the bifundamental matters. The situation is
different if we want to turn on mass deformation for anti-symmetric
matter, the background is an affine bundle. We will not discuss this
complication in this paper.

The Seiberg-Witten curve of the above quiver gauge theories without
mass deformation is
\begin{equation}
z^n+A(z)+\sum_{s=1}^r{B_s(z)+yC_s(z)\over
x-x_s}+\sum_{p=1}^q{yD_p(z)\over x-e_p}=0,~~k=2n,
\end{equation}
here $x_s$ are positions of NS5 branes which don't intersect with
the orientifold; $q$ is the number of NS branes which intersect with
the orientifold planes and $e_p$ are positions of $NS5$ branes stuck
at orientifold. This is natural since $e_p$ are fixed points under
the orbifold action. $A(z)$ and $B_s(z)$ are polynomials in z
\begin{equation}
A(z)=\sum_{l=1}^{n}A_ lz^{n-l},~~B_s(z)=\sum_{l=1}^{n}B_{sl}z^{n-l},
\end{equation}
and $C_s$ and $D_p$ are polynomials in z
\begin{equation}
C_s(z)=\sum_{l=2}^{n}C_{sl}z^{n-l}~~~D_p(z)=\sum_{l=2}^{n}D_{pl}z^{n-l}.
\end{equation}
We also have the constraint:
\begin{equation}
\sum_{s=1}^r C_s(z)+\sum_{p=1}^q D_p(z)=0.
\end{equation}
This curve can be derived by first write the Seiberg-Witten curve of
elliptic model, and then impose the orbifold invariance and finally
express it in terms of orbifold invariant variable. The
mass-deformed Seiberg-Witten curve is
\begin{equation}
z^n+A(z)+\sum_{s=1}^r{B_s(z)+yC_s(z)\over
x-x_s}+\sum_{p=1}^q{(y-y_p)D_p(z)\over x-e_p}=0
\end{equation}
where $y_p=\sqrt{Q(e_p)}$. $A(z), B(z), C(z), D(z)$ are polynomials
in $z$ of order $n-1$.

The Seiberg-Witten differential is given by
\begin{equation}
\lambda={ydx\over\prod_{i=1}^4(x-e_i)}.
\end{equation}

We will rewrite the above curve in a form along the way in
\cite{Gaiotto}. Let's first consider case ii) with two USp ends,
which corresponds to $q=0$. We rewrite the Seiberg-Witten curve in a
form which makes the interpretation with the $A_{2n-1}$ theory
compactification on a punctured sphere manifest. Expanding the
Seiberg-Witten curve in terms of polynomial of z, we have
\begin{equation}
z^n+\sum_{l=1}^{n }{p_r^l(x)\over
\Delta^{'}}z^{n-l}+\sum_{l=2}^{n}{yp^l_{(r-2)}(x)\over\Delta^{'}}z^{n-l}=0,
\end{equation}
here $\Delta^{'}=(x-x_1)....(x-x_r)$ and $p_r^l(x)$ are polynomials
with order $r$; $p^l_{(r-2)}$ are $r-2$ order polynomials. Define
$z=\prod_{i=1}^4(x-e_i)t^2$, then
\begin{equation}
y=t\prod_{i=1}^4(x-e_i)
\end{equation}

The Seiberg-Witten differential becomes
\begin{equation}
\lambda=tdx
\end{equation}
and the Seiberg-Witten curve is
\begin{equation}
t^{2n}+\sum_{l=1}^n{p_r^l(x)\over
\Delta^{'}\prod_{i=1}^4(x-e_i)^l}t^{2n-2l}+\sum_{l=2}^n{p_{r-2}^l(x)\over\Delta^{'}\prod_{i=1}^4(x-e_i)^{l-1}}t^{2n-2l+1}=0
\end{equation}
With this form, we conclude that this theory can be realized as the
six dimensional $A_{2n-1}$ theory compactified on a sphere with $r$
basic punctures $x_1,...x_r$ (see Figure 9a) for the Young Tableaux
and 4 generic punctures $e_i, i=1,..4$ with Young Tableaux in Figure
9b). The defects at the punctures are:
\begin{equation}
\phi_{2l}={p_r^l(x)\over
\Delta^{'}\prod_{i=1}^4(x-e_i)^l}dx^{2l},~~\phi_{2l-1}={p_{r-2}^l(x)\over\Delta^{'}\prod_{i=1}^4(x-e_i)^{l-1}}dx^{(2l-1)}.
\end{equation}

To clarify one point, $x$ is a coordinate on $C$, and since we
do not put any singularity at $\infty$, we can add a point at
$\infty$ to $C$ and compactify it to a sphere. This does not change
the Seiberg-Witten differential and other properties of our model.

Several checks can be made about this conclusion:

a)The moduli space of the sphere with $r+4$ punctures has dimension
$r+1$ which can be identified with the coupling constant of gauge
groups in the quiver.

b)The various differentials have pole $p_i=1$ at punctures at $x_s$
, which can be associated with the flavor symmetry $U(1)$, where the Young
Tableaux is shown in Figure 9a). The punctures $e_i$ has order
$p_l={l\over2}$ when $l$ is even, $p_l={{l-1}\over2}$ when $l$ is
odd. This puncture can be represented as a Young Tableaux with two
columns of height $n$ in Figure 9b). These poles correspond to
$SU(2)$ flavor symmetry. The total flavor symmetry is then
$SU(2)^4\times U(1)^r$, which matches the flavor symmetries read
from the quiver diagram.

c)For the differential $\phi_{2l}$, the dimension is
$4l+r-2(2l)+1=r+1$, which matches the dimension of the polynomials
$p_r^l$. For differential $\phi_{2l-1}$, the dimension is $r-1$,
which also matches the parameters needed for the polynomial
$p_{r-2}^{l}$.

d)When the mass deformation is turned on, we have the $t^{2n-1}$
term. Do a linear transformation on $t=t^{'}+\alpha$ to eliminate
this term. And keep the Seiberg-Witten differential as
$\lambda=t^{'}dx$. One can check the residue of the punctures $x_s$
and $e_p$ have the same patter as determined by the Young Tableaux.

Next, we consider case $i)$ for which only one NS5 brane intersects
with the O6 plane. The Seiberg-Witten curve is
\begin{equation}
z^n+A(z)+\sum_{s=1}^r{B_s(z)+yC_s(z)\over x-x_s}+{yD_1(z)\over
x-e_1}=0,~~k=2n
\end{equation}
Expand the curve in the polynomial of z and define
$z=\prod_{i=1}^4(x-e_i)t^2$, the curve becomes
\begin{equation}
t^{2n}+\sum_{l=1}^n{p_r^l(x)\over
\Delta^{'}\prod_{i=1}^4(x-e_i)^l}t^{2n-2l}+\sum_{l=2}^n{p_{r-1}^l(x)\over\Delta^{'}\prod_{i=2}^4(x-e_i)^{l-1}(x-e_1)^l}t^{2n-2l+1}=0
\end{equation}
Similarly, we conclude that this theory can be realized as the six
dimensional $A_{2n-1}$ compactified on a sphere with $r$ punctures
at $x_s$ and $3$ punctures at $e_i, i=2,3,4$, we also have a
different puncture at $e_1$ with Young Tableaux in Figure 9c). The
defects at the punctures are:
\begin{equation}
\phi_{2l}={p_r^l(x)\over
\Delta^{'}\prod_{i=1}^4(x-e_i)^l}dx^{2l},~~\phi_{2l-1}={p_{r-1}^l(x)\over\Delta^{'}\prod_{i=2}^4(x-e_i)^{l-1}(x-e_1)^l}dx^{(2l-1)}
\end{equation}
Similar checks can be made:

a)The dimension of moduli space of the punctured sphere is $r+1$
which is identified with the $r+1$ coupling constants of gauge
groups.

b)The flavor symmetries correspond to $x_s$ are U(1), while $e_i,
i=2,3,4$ represent flavor symmetry $SU(2)$. The $e_1$ puncture has
pole $p_{l}={l\over2}$ for $l$ even, and $p_{l}={(l+1)\over2}$ for
odd $l$. This can be represented by the Young Tableaux in Figure
9c). The flavor symmetry of this puncture is $U(1)$. Therefore, the
total flavor symmetry is $U(1)^r\times SU(2)^3\times U(1)$, Which
matches our counting from the quiver diagram. Note that the Young
Tableaux for the $U(1)$ from the two fundamentals on the SU ends is
different from the $U(1)$ punctures for the bi-fundamental matter.

c)The dimension of $\phi_{2l}$ is $r+1$, and $\phi_{2l-1}$ has
dimension $r$, which matches the parameters needed for the
polynomial $p_r^l(x)$ and $p_{r-1}^l(x)$.

d)The flavor symmetry can be checked from the mass deformed theory.

\begin{figure}
\begin{center}
\includegraphics[width=4in,]
{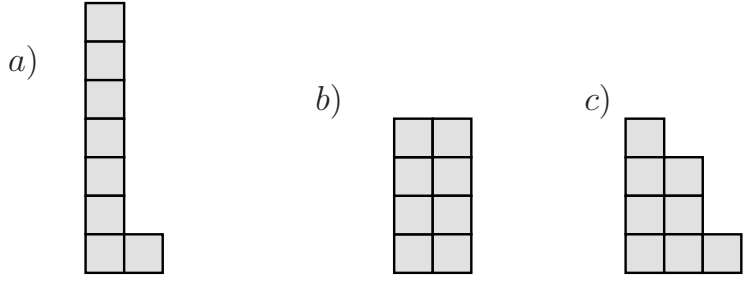}
\end{center}
\caption{Young-Tableaux for a): Puncture with $p_i=1$ b): Puncture
with $p_l={l\over2}$ for even l, $p_l={(l-1)\over2}$ for odd l; c):
Puncture with $p_l={l\over2}$ for even l, $p_l={(l+1)\over2}$ for
odd l.}
\end{figure}

Finally, let's consider the quiver in case $iii)$; the
Seiberg-Witten curve can be written as
\begin{equation}
t^{2n}+\sum_{l=1}^n{p_r^l(x)\over
\Delta^{'}\prod_{i=1}^4(x-e_i)^l}t^{2n-2l}+\sum_{l=2}^n{p_{r-1}^l(x)\over\Delta^{'}\prod_{i=3}^4(x-e_i)^{l-1}(x-e_1)^l(x-e_2)^l}t^{2n-2l+1}=0
\end{equation}

Similarly, this theory can be written as the six dimensional
$A_{2n-1}$ theory compactified on Riemann surface with punctures
$e_i$ and $x_s$. The defects at the punctures are:
\begin{equation}
\phi_{2l}={p_r^l(x)\over
\Delta^{'}\prod_{i=1}^4(x-e_i)^l}dx^{2l},~~\phi_{2l-1}={p_{r-1}^l(x)\over\Delta^{'}\prod_{i=3}^4(x-e_i)^{l-1}(x-e_1)^l(x-e_2)^l}dx^{(2l-1)}
\end{equation}
One can check along the similar line that this is the correct
interpretation.

\section{Some Special Examples}
We want to mention some special examples which are of later interest
for us.  We first analyze $SU(2n)$ with two-antisymmetric matter and
four fundamentals, this corresponds to $r=0, q=2$. The
Seiberg-Witten curve is
\begin{equation}
0=t^{2n}+\sum_{l=1}^n{A_l\over\sum_{i=1}^4(x-e_i)^l}t^{2n-2l}+\sum_{l=2}^n{D_l\over
\sum_{i=3}^4(x-e_i)^{l-1}(x-e_1)^l(x-e_2)^l}t^{2n-2l+1}
\end{equation}
So this theory can be represented as the $A_{2n-1}$ theory
compactified on a sphere with four punctures, two of which have the
form as Figure 9a, and two of which have the form as Figure 9b.

We then study the quiver gauge theory corresponding to $r=1, q=0$,
the quiver gauge theory is $USp(2n)\times USp(2n)$. The flavor
symmetry in this case is $SU(2)^4\times SU(2)$. The last $SU(2)$
comes from the bifundamental matter which now furnish a real
representation of quiver theory. Naively, we identify this theory as
$A_{2n-1}$ compactified on a sphere with four punctures $e_i$ and
one basic puncture $x_1$. The manifest flavor symmetry from this
representation is $SU(2)^4\times U(1)$.

Finally, we consider the quiver corresponding to $r=0, q=1$, this is
a $USp(2n)$ theory with four fundamental and one-antisymmetric
hypermultiplet. The Seiberg-Witten curve is
\begin{equation}
t^{2n}+\sum_l{p_l\over \prod_{i=1}^4(x-e_i)^l}t^{2n-2l}=0
\end{equation}
This theory is represented as $A_{2n-1}$ theory compactified on
sphere with four identical puncture with $SU(2)$ flavor symmetry.
Combined with the permutation symmetry of this four identical
punctures, we expect that this theory has the $SL(2,Z)$ duality.
Notice that the above curve can be written as
\begin{equation}
(t^2+{q\over\prod_{i=1}^4(x-e_i)})^n=0.
\end{equation}
It is amusing to note that for $SU(2)$ theory with four
foundamentals, the Seiberg Witten curve is
\begin{equation}
t^2+{q\over\prod_{i=1}^4(x-x_i)}=0.
\end{equation}
So the Seiberg-Witten curve for $USp(2n)$ theory with four
fundamentals and one traceless anti-symmetric representation is
tensor product of that of $SU(2)$ theory with four fundamentals
\cite{schwarz}.

\section{Degeneration Limit}
In reference \cite{Gaiotto}, it is conjectured that the SU quiver
gauge theory $SU(2)\times SU(4)\times SU(6)...\times
SU(2n)^{m-2n+4}\times...SU(4)\times SU(2)$ is S dual to quiver gauge
theory $SU(2)\times SU(3)\times SU(4)\times...SU(2n)^{m-2n+3}\times
USp(2M)$. The SU quiver gauge theory is realized as the
compactification of $A_{2n-1}$ theory compactified on a sphere with
$m+1$ basic punctures and two special puncture with $SU(2)$ flavor
symmetry. When two special punctures collide, a USp gauge group is
decoupled. More generally, when both ends are associated with USp
group, it is related to a sphere with four $SU(2)$ punctures and
several basic punctures.
\begin{figure}
\begin{center}
\includegraphics[width=4in,]
{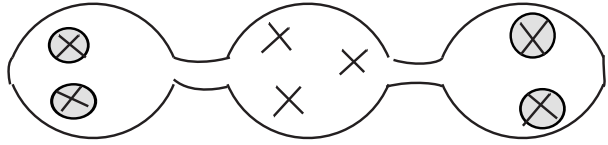}
\end{center}
\caption{The degeneration limit corresponds to two weakly coupled
USp group, two $SU(2)$ punctures are colliding.}
\end{figure}

This conjecture is proved in this paper by rewriting the
Seiberg-Witten curve of $USp\times SU^{r-1}\times USp$ quiver in a
form which makes the above interpretation manifest. We show that the
quiver with two USp ends are associated with the sphere with several
basic punctures and four $SU(2)$ punctures. When two $SU(2)$
punctures collide with each other, the gauge coupling of the USp group
becomes weakly coupled. The linear quiver with two USp ends
associated with the degeneration limit is shown in Figure 10.  When we turn
off one of weakly coupled gauge coupling, we are left with a puncture
associated with the flavor symmetry $SU(2n)$, which can be seen from
the linear quiver.

The theory with only one USp node can be derived by colliding
several basic punctures. It is associated with a sphere with two
$SU(2)$ punctures and several basic punctures, see Figure 11a). The
Young tableaux associated with $SU(2)$ flavor symmetry implies the
tail $n_1=2, n_2=4,..n_k=2k, n_n=2n$, if we collide a basic puncture
with a $SU(2)$ puncture, a $SU(2)$ gauge group becomes weakly
coupled, we are in another corner of moduli space around which we
have the weakly coupled description. In this description, the quiver
becomes $SU(2)\times SU(4)\times SU(6)...\times
SU(2n)^{m-2n+4}\times...SU(4)\times SU(2)$. The degeneration limit
of the punctured sphere corresponding to two quiver are shown in
figure 11b).
\begin{figure}
\begin{center}
\includegraphics[width=4in,]
{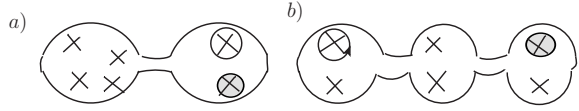}
\end{center}
\caption{a) Weakly coupled description with USp ends; b) Weakly
coupled description with all SU chain groups with bifundamental and
fundamental matters.}
\end{figure}

The $T_N$ theory can be derived by using a sphere with four $SU(2)$
punctures and $N-1$ basic punctures. Figure 12 shows how we can get
the $T_N$ (N must be even) theory by colliding the punctures.
\begin{figure}
\begin{center}
\includegraphics[width=4in,]
{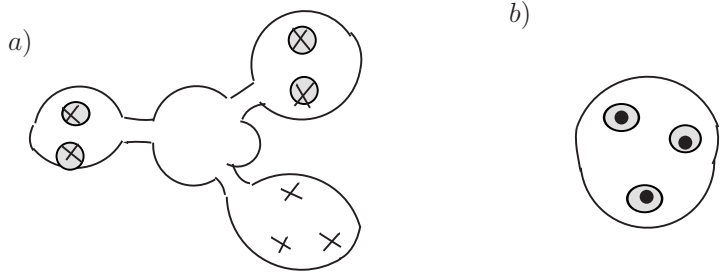}
\end{center}
\caption{a) One degeneration limit of quiver gauge theory with USp
ends and $N-2$ SU gauge groups, here $N=4$; b)$T_N$ theory when we
turn off the weak gauge couplings completely. }
\end{figure}

We can also study the degeneration limit of quiver theory which has
$SU$ ends with antisymmetric matter. The linear quiver with two such
SU ends is depicted as the six dimensional theory compactified on a
sphere with several punctures in Figure 13.
\begin{figure}
\begin{center}
\includegraphics[width=4in,]
{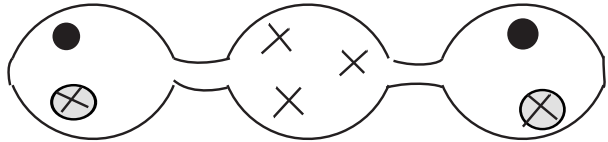}
\end{center}
\caption{The degeneration limit corresponds to two weakly coupled SU
group with antisymmetric representation, one SU(2) puncture and a
special $U(1)$ puncture are colliding.}
\end{figure}
We can similarly conclude that the linear quiver with one SU ends
with antisymmetric matter is associated with a sphere with several
basic puncture, one special puncture and a SU(2) puncture. This can
be shown by colliding several basic punctures of above theory with
two special $SU$ ends. When special $U(1)$ puncture collides with
$SU(2)$ puncture, a $SU(2n)$ gauge theory becomes weakly coupled.
When we turn off this coupling completely, we are left with a
$SU(2n)$ puncture, which can be seen from our linear quiver.

Motivated by the discussion of USp theory,  We can conclude from the
form of two special punctures that this quiver is S dual to linear
quiver with two tails, one tail associated with the Young Tableaux of
special U(1), the other associated with $SU(2)$. For the special
$U(1)$, the Young-Tableaux implies that the tail has the form
$n_1=3, n_2=5,..n_k={2k+1},..n_{n-1}=2n-1,n_{n}=2n$, while the other
tail has the form $n_1=2, n_2=4,..n_k=2k, n_n=2n$. Two different
degeneration limits are depicted in figure 14. We also write the two
different  linear quivers associated with these punctured spheres in
Figure 15 with $2n=6$.
\begin{figure}
\begin{center}
\includegraphics[width=4in,]
{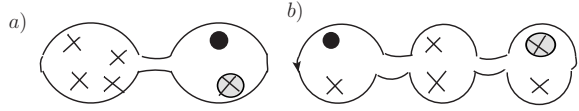}
\end{center}
\caption{a) Linear quiver with weakly coupled $SU$ ends with
antisymmetric representation, b) S dual quiver to theory $a$ with SU
chain of gauge groups with bi-fundamental and fundamental matters.}
\end{figure}
\begin{figure}
\begin{center}
\includegraphics[width=4in,]
{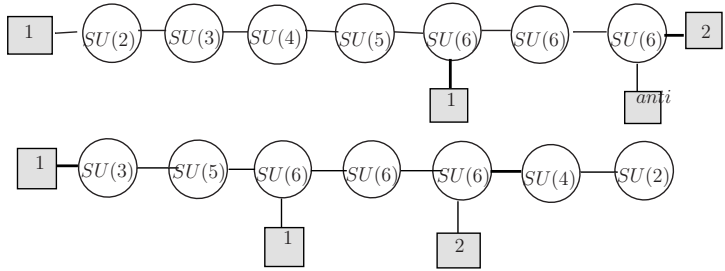}
\end{center}
\caption{a)Linear quiver with weakly coupled $SU$ ends with
antisymmetric representation, b)S dual to $a$ with SU chain of gauge
groups}
\end{figure}
Similarly, we can find $T_{2n}$ theory from $SU(2n)\times
SU^{n-2}\times SU(2n)$, see figure 16. We can also find $T_{2n}$
from theory with form $SU(2n)\times SU^(n-2)\times USp(2n)$ by
colliding appropriate punctures.
\begin{figure}
\begin{center}
\includegraphics[width=4in,]
{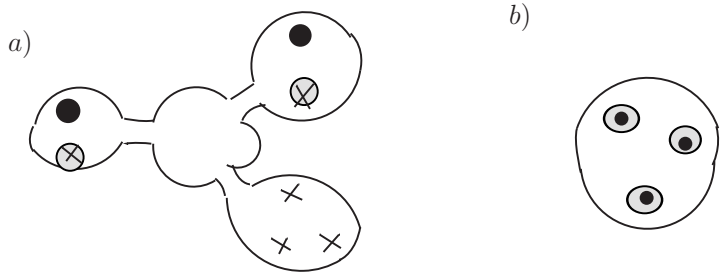}
\end{center}
\caption{$T_{2n}$ theory from $SU(2n)\times SU^{n-2}\times SU(2n)$
quiver. }
\end{figure}

Next, let's have some fun with other isolated superconformal field
theory, as in the examples outlined in \cite{argy3}. First, let's
consider $SU(4)$ theory with four fundamentals and two antisymmetric
hypermultiplets. The weakly coupled $SU(4)$ description is depicted
in Figure 17a). It is associated with $A_3$ theory compactified on a
sphere with two $SU(2)$ punctures and two special $U(1)$ punctures.
There is another degeneration limit, which is depicted in Figure
13b). As we discussed earlier, when two $SU(2)$ punctures collide, a
$USp(4)$ group becomes weakly coupled, if we turn off this gauge
coupling, we are left with a $SU(4)$ puncture. The resulting three
punctured sphere is shown in Figure 13c).
\begin{figure}
\begin{center}
\includegraphics[width=4in,]
{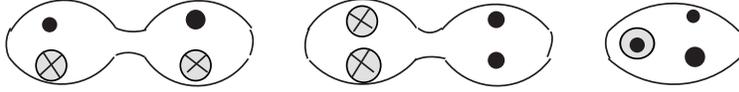}
\end{center}
\caption{a) Theory with weakly coupled SU group; b) Theory with
weakly coupled USp group; c)$E_6$ theory.}
\end{figure}

For the three punctured spheres, the order of poles at each puncture
are
\begin{equation}
SU(4)~puncture~: p_2=1, p_3=2, p_4=3;
\end{equation}
\begin{equation}
Special~U(1)~puncture~:p_2=1,p_3=2,p_4=2.
\end{equation}

The Seiberg-Witten curve for the tree punctured sphere can be read
from our rules (we put three punctures at $x=1,0,\infty$:
\begin{equation}
t^4+{U\over (x-1)^2x^2}t=0
\end{equation}
One can check that the other differentials are vanishing by using
formula $(2.5)$. This theory has a dimension 3 operator. The curve
reduces to
\begin{equation}
t^3+{U\over (x-1)^2t^2}=0
\end{equation}
which is exactly same as the $E_6$ superconformal field theory. The
manifest flavor symmetry is $SU(4)\times U(1)\times U(1)$, which is
a subgroup of $E_6$. We expect the flavor symmetry is enhanced to
$E_6$ and we identify this theory as $E_6$ superconformal field
theory.

More generally, if we consider $SU(2n)$ theory, then the theory
associated with three punctured sphere has the manifest flavor
symmetry $SU(2n)\times U(1)\times U(1)$. The original theory has
operators $D(\phi)=2,3...2n$. In the dual description, a $USp(2n)$
group is becoming weakly coupled, and it has even dimensional
operators. Therefore, this isolated superconformal field theory only
has odd dimensional operators with $D(\phi)\geq3$, its
Seiberg-Witten curve can be written from the puncture types
according to our general rule.

Then let's consider the quiver $USp(4)\times USp(4)$ which has been
described in section 4. This theory is associated with $A_3$ theory
compactified on a sphere with four $SU(2)$ puncture and a $U(1)$
puncture. These two weakly coupled USp group description correspond
to the degeneration limit shown in Figure 18a). There is another
degeneration limit shown in Figure 18b). Here two $SU(2)$ punctures
collide and one $SU(2)$ puncture collide with other $U(1)$ punctures.
When two $SU(2)$ puncture collide, a $USp(4)$ gauge group becomes
weakly coupled. When it is completely turned off, a $SU(4)$ puncture
is left. On the other hand, when a $SU(2)$ puncture collides with a
basic puncture, a $SU(2)$ gauge group becomes weakly coupled. When
we turn off this weakly coupled gauge group, a puncture with Young
Tableaux of only one row $n_1=4$ is created, this is associated with
a $SU(4)$ puncture.

The resulting three punctured sphere has two $SU(4)$ punctures and
one $SU(2)$ puncture. This theory has only one dimension 4 operator.
The original quiver has two dimension 2 and two dimension 4
operators; in the dual description, we have a dimension 2 operator
for $SU(2)$ group and a two dimensional and four dimensional
operators for $USp(4)$ group, so we are left a dimension 4 operator
for this isolated superconformal field theory.

The $A_3$ theory compactified on such punctured sphere is identified
as $E_7$ superconformal field theory \cite{Gaiotto}\cite{Argy}. Note
that the manifest flavor symmetry is $SU(4)\times SU(4)\times
SU(2)$, which is a maximal subgroup of $E_7$.

More generally, when we consider $USp(2n)\times USp(2n)$ theory, the
three punctured spheres have flavor symmetry $SU(2)\times SU(2) \times
SU(2n)\times SU(2)$. For generic $n$, the collision of basic
puncture and $SU(2)$ puncture creates a puncture with flavor
symmetry $SU(2)\times SU(2)$. This theory has only even dimension
operators $D(\phi)\geq4$.
\begin{figure}
\begin{center}
\includegraphics[width=4in,]
{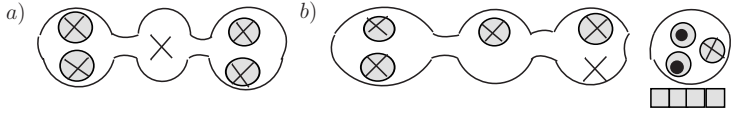}
\end{center}
\caption{a)Theory with weakly coupled SU group b) Theory with weakly
coupled USp group, c)$E_6$ theory.}
\end{figure}

\section{Conclusion}
In this paper, we have studied $N=2$ linear $SU$ quiver gauge theory with
USp ends or SU ends with antisymmetric representations. We rewrite
the Seiberg-Witten curve in a form which makes manifest the
interpretation of six dimensional $A_N$ theory compactification on
punctured sphere. We identified the flavor symmetry of the theory with
the punctures. We then study the degeneration limit of those
theories and identify the weakly coupled description in various
cusps of moduli space. For the USp ends, we check the previous
conjecture; For SU ends, we conjecture a dual quiver with ordinary
SU chain with bi-fundamental and fundamental matters. Finally, we have
seen how $E_6$ and $E_7$ superconformal field theories come from the
degeneration limit of certain special field theories. We also found
a class of isolated superconformal field theories with odd dimension
operators starting from dimension 3 and superconformal field theory
with even dimension operators starting from dimension 4.

We only considered massless antisymmetric matter in this paper. It
would be interesting to study the mass deformed theory and identify
the six dimensional description. The mass deformed antisymmetric
matter theory changes the background from a product manifold
$C\times T^2$ to an affine bundle , which is similar to the most
generic mass deformed elliptic model. The addition of mass
deformation of antisymmetric matter may change the picture
dramatically. We can see this from the elliptic model with only one
gauge group. Without mass deformation, the four dimensional theory
is defined as the six dimensional $(0,2)$ theory compactified on a
smooth torus, and we have $N=4$ supersymmetry. Now if we turn on the
mass deformation for the adjoint hypermultiplet, this theory is
described by a torus with one puncture, and we only have $N=2$
supersymmetry.

It is also interesting to study the quiver with $SO$ node and $SU$
node with symmetric representation. The Type IIA brane construction
\cite{uranga} involves a orientifold six plane with positive charge
and a negative charged orientifold six plane. It would be interesting to
find a six dimensional description.
\begin{flushleft}
\textbf{Acknowledgments}
\end{flushleft}
We are grateful to Eric Mayes for carefully reading the manuscript.
This research was supported in part by the Mitchell-Heep chair in
High Energy Physics (CMC), by the Cambridge-Mitchell Collaboration
in Theoretic Cosmology, and by the DOE grant DEFG03-95-Er-40917.

\end{document}